\documentclass[11pt]{article}
     \usepackage[margin=0.9in]{geometry}
     \usepackage{times}
     \usepackage{lipsum}
\usepackage{setspace}
\usepackage{verbatim}

\usepackage{titling}
\usepackage{sectsty}

\usepackage{amsmath}
\usepackage{titling}

\usepackage [english]{babel}
\usepackage{enumitem}

\usepackage{etex}
\usepackage{amsthm}
\usepackage{amssymb}
\usepackage{soul}
\usepackage{calc}
\usepackage{newtxtext,newtxmath}

\usepackage{colortbl}
\usepackage[boxed]{algorithm2e}
\usepackage{epstopdf}
\usepackage{arydshln}
\usepackage{booktabs}
\usepackage{natbib}
\usepackage{relsize} 

\usepackage{ragged2e}
\justifying

\usepackage[usenames,dvipsnames]{xcolor}
\usepackage{hyperref}
\hypersetup{
	colorlinks=true,         
	linkcolor=MidnightBlue,          
	citecolor=MidnightBlue,
	urlcolor=MidnightBlue            
 }

\usepackage{xcolor} 

\usepackage{breakurl}
\usepackage{bookmark}
\usepackage{etoolbox}
\usepackage{graphicx}
\theoremstyle{plain}
\theoremstyle{definition}

\newtheorem{proposition}{Proposition}
\usepackage[title]{appendix}
\usepackage{array}

\usepackage{authblk}
\newcommand{\be}{\begin{equation}}
\newcommand{\ee}{\end{equation}}

\newcommand{\subtitle}[1]{%
  \posttitle{%
    \par\end{center}
    \vspace{-2em}
    \begin{center}\Large#1\end{center}
    \vskip0.5em}%
}

\title{Relational Observables and Physical Perspectivalism in General Relativity: The View from Nowhere vs. The View from Everywhere}

\author[1,2]{Nicola Bamonti\textsuperscript{}\thanks{\href{mailto:nicola.bamonti@sns.it}{nicbamonti@gmail.com}}}

\date{ }

\affil[1]{Department of Philosophy, Scuola Normale Superiore, Piazza dei Cavalieri, 7, Pisa, 56126, Italy}
\affil[2]{Department of Philosophy, University of Geneva, 5 rue de Candolle, 1211 Geneva 4, Switzerland}

\begin{document}

\maketitle

\vskip .8cm \centerline{\bf Abstract} \vskip 0.2cm 
In General Relativity, relational constructions yield gauge-invariant Dirac observables by expressing physical fields relative to reference frames. This leaves open how the physical situations represented in different frames should be understood. I articulate two ontological interpretations within a common fibre-bundle framework, distinguishing \emph{frame-independence} from \emph{frame-freedom} to clarify appeals to perspective-neutrality. 
The \emph{View from Nowhere} treats the situations characterised by relational observables as partial aspects of a shared physical situation represented by a frame-free gauge-equivalence class. 
The \emph{View from Everywhere} takes each relational observable to characterise a comprehensive physical situation in its own right, without commitment to a shared frame-free reality. They articulate a qualified realisation of moderate and strong physical perspectivalism within GR, respectively. 
Using two GPS reference frames as a working example, I show that a physically constrained, frame-independent inter-frame map supplies a constructive counterexample to the claim that connecting perspectives requires frame-free structures in the ontology. 
The choice between the two interpretations remains open, with their ontological commitments made explicit. I conclude by outlining possible implications for quantum reference frames and relational quantum mechanics.


\section{Introduction}\label{Intro}

Discussions in foundations of physics on what count as physically meaningful descriptions have increasingly drawn on relational and gauge-theoretic frameworks, especially in connection with quantum reference frames and relational observables. Physical descriptions are often understood as formulated relative to reference frames or observers, rather than as absolute descriptions detached from any perspective. Applied to General Relativity (GR), this raises a precise interpretive question: \emph{how should we interpret two distinct relational observables defined relative to two distinct reference frames?} Are they partial accounts of one shared physical situation, or does each characterise a physical situation that is comprehensive in its own right?

In this paper, I argue that both readings can be articulated precisely within GR and the case of two materially distinct reference frames brings the issue into focus. The issue concerns both what counts as a physical situation and whether connections between frame-relative descriptions require a further physical situation common to them. The relational observables framework makes it possible to examine these questions within a clear, formal vocabulary in which a set of central concepts, such as `observer', `perspective', `objective', and  `perspective-neutrality' acquire precise meanings, thus contributing to debates on \textit{physical perspectivalism} in gauge theories. 

In gauge-theoretic frameworks such as GR, following Dirac's famous prescription, an \emph{observable} is a quantity invariant under the theory’s gauge symmetries \citep{Dirac:1964}: transformations that leave the physical content of the theory, also referred to as the \lq{}observable content\rq{}, unchanged.\footnote{\label{fn:dirac} In canonical terms: a phase-space function $O$ is a Dirac observable iff $\{O,C_\alpha\}\approx0$ for all first-class constraints $C_\alpha$, equivalently $O$ is constant along the gauge orbits they generate. For an overview of representational redundancy across physics and philosophy, see \citet{Berghofer2023-BERGS-9}.} 
On this understanding, gauge-related models are different mathematical representations of the same physical situation, and only gauge-invariant quantities, also known as Dirac observables, carry physical content.

GR places a distinctive strain on this picture. The relevant gauge group is the diffeomorphism group, whose action \lq{}reshuffles\rq{} the manifold points on which fields are defined, typically as sections of some vector bundle over $\mathcal{M}$. Hence any quantity \textit{locally} defined at a manifold point, or even as an integral over a finite spacetime region without further structure, fails to be gauge-invariant. The immediate consequence is familiar: the natural candidates for \textit{local} observables fail to qualify as observables in the Dirac sense. 
This has often been taken to suggest that only global observables are available. However, global observables form too small a class: they do not provide a \textit{complete set of observables} that would uniquely characterise the physical situation \citep{Komar,Torre1993}. 

Relational constructions address this tension by locating observables  \emph{relationally}, relative to the values of some additional set of physical fields acting as a \emph{reference frame}, rather than anchoring it to bare manifold points \citep{Tambornino2012}. \emph{Reference frames} in GR may be built from intrinsic curvature scalars \citep{Komar,Kretschmann1917} or material fields \citep{DeWitt1962}; they play a crucial role in the partial observable framework \citet{Rovelli_2002,Dittrich2006,Dittrich2007} and underpin the construction of  \textit{relational observables} \citep{Bamonti2023,Bamonti2026}.

As the central interpretive question of the paper concerns the interpretation of relational observables, it is necessary to provide some technical precision about their construction from the outset.

A typical reference frame will be defined here as a set of four scalar degrees of freedom  $\Phi:=\{\phi^{(I)}\}|_{I=1,\dots,4}$ associated with a physical (typically, but not necessarily material) system satisfying some dynamical equations and that establish a \textit{local} diffeomorphism\footnote{Global reference frames are not usually available due to topological obstructions similar to Gribov obstructions \citep{Gribov1978}. Even when the topology allows it, reference frames may generically develop caustics due to their dynamical nature. \label{fngribov}}
\begin{equation}
    \Phi: U \subset \mathcal{M} \longrightarrow V \subset \mathbb{R}^{4},\qquad
    p \longmapsto \big(\phi^{(1)}(p),\phi^{(2)}(p),\phi^{(3)}(p),\phi^{(4)}(p)\big).\label{mapframe}
\end{equation} 
A minimal \textit{viability} requirement for such a mapping is local invertibility:
\begin{equation}
    \det\!\left(\frac{\partial \phi^{(I)}}{\partial x^{\mu}}\right)\neq 0
    \quad \text{in some open region } U\subset\mathcal{M},\label{jacobian}
\end{equation}
so that the local inverse $\Phi^{-1}$ then allows any field quantity $T_{[ab\dots]}$ to be re-expressed \textit{relationally} as
\begin{equation}
    T_{[IJ\dots]}(\Phi):=[(\Phi^{-1})^{*}]T_{[ab\dots]},\label{pullback}
\end{equation}
where $[\bullet]^{*}$ denotes the pullback by $\Phi^{-1}$  and $T_{[ab\dots]}$ denotes a covariant tensor understood locally as a section of some appropriate tangent vector bundle over $\mathcal{M}$.\footnote{Here, I am using the abstract index notation (see \citealp{Penrose1984}) to stress that it is a geometrical object independent from the choice of a coordinate representation.}
This is what is often referred to as a \textit{relational observable}.

The construction secures the relational observables, but does not settle the interpretive question posed above.

The answer bears on the viability of two forms of \emph{physical perspectivalism}, moderate and strong \citep{AdlamPerspect}. To better articulate these metaphysical positions within GR, in §\ref{terminology} we distinguish between two properties a quantity may possess: \textit{frame-independence}---the preservation of its value across changes of frame---and \textit{frame-freedom}---its construction without invoking any frame. 

Within this regimented vocabulary, two distinct interpretive ontological options can be cleanly separated. 

The \textit{View from Nowhere} (\S\ref{VFN}) treats relational observables as gauge-invariant perspectival representations of an underlying shared physical situation, formalised as an equivalence class under the gauge group. This View gives a realisation of Adlam's \textit{moderate} physical perspectivalism, with perspective-neutrality understood as frame-freedom and typically informs frameworks such as the \textit{perspective-neutral} approach as presented in \cite{Vanrietvelde2020} and related work on quantum reference frames \citep[e.g.][]{delaHametteperspective,QuantumHole}.\footnote{Explicitly, \cite{QuantumHole} state \lq{}\lq{}[\dots] changes of quantum reference frame [\dots] \textit{should not affect the physical situation}. Thus, whether a system is in superposition or is entangled, and how these properties change dynamically, becomes a mere matter of perspective [emphasis added]\rq{}\rq{}. A different approach can be found in \cite{Giacomini2019}.}
 
The \textit{View from Everywhere} (\S\ref{sec3}) takes \textit{each} relational observable to characterise a physical situation that is ontologically comprehensive in its own right. It rejects commitment to a shared frame-free physical situation while retaining physically constrained relations between the frame-relative situations. On the frame-free reading of perspective-neutrality, this articulates a form of strong physical perspectivalism within GR. 

The paper does not settle the choice between these Views.  It distinguishes what each relational description encodes from the physical situation it is taken to represent, making their competing commitments explicit. For this purpose, I adopt the \emph{fibre-bundle formalism}, which provides a common construction of gauge-equivalence classes, relational observables and inter-frame maps within which the two Views can be compared and regimented in GR and, possibly, in gauge-theories in general \citep{gomesgaugetheorygeometry,GomesBook,gomes2026geometry}. 

\paragraph{Roadmap.}
The argument begins by distinguishing frame-dependence, frame-independence and frame-freedom within a minimal relational vocabulary (\S\ref{terminology}), and by constructing relational observables through Komar's curvature-scalar frame within the fibre-bundle formalism (\S\ref{sec1}). This supplies the tools for distinguishing the senses of perspective-neutrality at stake in the physical perspectivalism debate (\S\ref{debunking}). Using two distinct GPS frames, \S\ref{VFN} develops the View from Nowhere's interpretation of fibre-bundle structures. \S\ref{sec3} uses the same construction to articulate the View from Everywhere and examine whether inter-perspectival connections require a frame-free ontology. \S\ref{conclusion} draws out the consequences and formulates questions for quantum reference frames and relational quantum mechanics.

\section{A Minimal Relationalist’s Glossary}\label{terminology}

The aim of this section is to set out a minimal glossary to classify the field-based quantities relevant to the argument, distinguishing relational from non-relational quantities and, subsequently, frame-independence from frame-freedom. This is not meant to be a complete taxonomy for \textit{all} geometrical structures relevant in GR. 

Importantly, the term ‘spatiotemporal’ will be reserved for quantities constructed using tensorial objects defined on the manifold as sections of some vector bundle over $\mathcal{
M}$, without committing to any interpretation of what represents spacetime.\footnote{The concept of spacetime lacks a universally agreed definition, and its interpretation varies across different frameworks. It may be understood as: (i) the manifold $\mathcal{M}$; (ii) the combination $(\mathcal{M}, g_{ab})$ of the manifold equipped with the metric field representing the gravitational field; or (iii) the gravitational field $g_{ab}$ alone \citep{Rovelli2021}, treating $\mathcal{M}$ as a purely mathematical construct without ontological status.}

\subsection*{Non-relational quantities}
Within the domain of \textit{non-relational quantities}, I distinguish the following categories:

\begin{enumerate}
\item[(i)] \textbf{Spatiotemporally explicit quantities:} tensorial objects defined intrinsically on the manifold as sections of some vector bundle over $\mathcal{
M}$. They can be distinguished into:
\begin{enumerate}
\item[(i.a)] \textbf{Spatiotemporally \textit{local}:}  Determined solely by the properties at a single \textit{point}, such as the local metric tensor $g_{ab}(p)$. 
\item[(i.b)] \textbf{Spatiotemporally \textit{non-local} (quasi-local or global):} I use \textit{quasi-local} for integrals over \textit{finite} spacelike or compact spacetime regions and along worldlines, such as proper time. \textit{Global} quantities concern integrals over the whole spacetime or an entire hypersurface, including ADM charges.
This classification concerns the extent of the integration domain.\footnote{This is not the only possibility in the literature. For instance, \citet{Torre1993}'s \textit{local} quantities are integral functionals having integrands depending on canonical fields and finitely many derivatives at a point. ADM charges are local in this sense, but global as geometric quantities: they are surface integrals at spatial infinity on an open, asymptotically flat hypersurface $\Sigma^3$.}
\end{enumerate}
\item[(ii)] \textbf{Spatiotemporally implicit quantities:} Not representable by a single object over $\mathcal{M}$, but only by a collection of spatiotemporally-explicit (typically local) objects---typically an equivalence class (hence the name \textit{implicit}). Such quantities are conventionally denoted using square brackets, e.g $[g_{ab}]$ represent the equivalence class of diff-related local metrics $g_{ab}(p)$.
\end{enumerate} 

\subsection*{Relational quantities}
There are various ways to define relational quantities. A quantity is \textit{relational} here if it is constructed through the frame-based dressing in \eqref{pullback}. The construction is diff-invariant under simultaneous transformations of field and frame, which play distinct roles: the object dressed and the structure localising it.\footnote{These roles need not involve independent fields: in \S\ref{sec1}, $g_{IJ}(\Re_g)$ is constructed by dressing the metric with its own Kretschmann--Komar scalars.} The definition also includes integral quantities, such as volumes defined by a relational metric over dressed regions. Merely comparing quantities does not therefore make them relational in this sense. For example, $R(p)-R(q)$,  with $R$ a scalar field and $p,q$ fixed manifold points, is not generally diff-invariant and is not relational in this sense. This is a bold choice, but it is purely instrumental to the main topic of this work. 

Given eq. \eqref{mapframe}, relational quantities can be categorised in a manner mirroring the manifold-based quantities discussed above, as follows:

\begin{enumerate}
\item[(I)] \textbf{Frame-explicit (or frame-relative) quantities:} objects defined on the dressed manifold $\Phi(\mathcal{M})\subseteq\mathbb{R}^4$, the frame's value space, as sections of vector bundles induced by the pullback $[\Phi^{-1}]^*$. For example, $g_{IJ}\in\Gamma(T^*\Phi(\mathcal{M})\otimes T^*\Phi(\mathcal{M}))$. These quantities may be: 
\begin{enumerate}
\item[(I.A)] \textbf{Frame (or relationally) \textit{local}:}  Determined solely by the properties at a dressed point $p:=\phi^{(I)^{-1}}(x)$, such as the relational metric $g_{IJ}(\Phi)$.\footnote{Relational locality does not imply locality in the sense of \citet{Torre1993}: frame dynamics typically make these observables non-local functionals of Cauchy data. Some gauge-fixing prescriptions also require elliptic reconstruction from spatial boundary data \citep{Wallace2024,Bamonti2026}.} These quantities are gauge-invariant only if the frame and the field transform \emph{together} under diffeomorphisms, a property which fails in the case of \textit{dynamical uncoupling} of the frame \citep{BamontiGomes2024,Bamonti2026}.
\item[(I.B)] \textbf{Frame \textit{non-local}  (quasi-local or global):}  The quasi-local/global distinction applies to regions dressed by the chosen frame. Global integrals over $\Phi(\mathcal{M})$ face the obstructions to global frames noted in footnote~\ref{fngribov}.

\end{enumerate}
\item[(II)] \textbf{Frame implicit quantities:} Quantities represented only by collections of frame-explicit objects, not necessarily equivalence classes. I denote them by $\{g_{IJ}\}$; the relations between their members are specified in \S\S\ref{VFN},\ref{sec3}.
\end{enumerate}

\subsection*{Dependence, independence, freedom}

Finally, building on \cite{Wallace2019} distinction between \textit{coordinate} \textit{independence} and \textit{freedom}, it is useful to separate three notions:

\begin{enumerate}

        \item \textbf{Frame-dependence:} A relational quantity is frame-dependent if both its functional form and the values it takes change when the frame changes.
    \item \textbf{Frame-independence:} A relational quantity is frame-independent if its functional form depends on the choice of the reference frame, but the values it takes \textit{at the dressed point} remains unaffected by the choice of reference frame in which it is formulated. Applying this criterion to maps between frames requires distinguishing
the two field-theoretic roles involved, as explained above.
    \item \textbf{Frame-freedom:} A quantity is frame-free if its construction does not adopt a reference frame.  Frame fields may occur among its constituents without being used in this role. Non-relational quantities exemplify frame-free quantities. 
\end{enumerate}

Let's be explicit on the well-known and crucial distinction between the \lq{}functional form\rq{} and \lq{}value\rq{} of a quantity.
According to the above-presented terminology, a relationally constructed scalar $R(\Phi):=[\Phi^{-1}]^*R$ is frame-\textit{in}dependent: as it is standard in covariant formalism, \textit{its functional form depends on the frame, but its value does not}.\footnote{Of course, being a scalar is not sufficient for frame-independence: the energy density $\rho:=T^{\rm stress-energy}_{ab}u^{a}u^{b}$ is a counterexample, as it is constructed explicitly with the four-velocity $u^a$ representing the congruence of observers that constitute the frame itself. As such, $\rho$ irreducibly depends on which physical system serves as the frame.}

A simple example makes the functional-form/value distinction explicit.\footnote{The frame clocks below share the FLRW foliation, so their change of frame reduces to a value-space relabelling (cf. \citealp{Brown1995} on dust symmetries). The example therefore illustrates the functional-form/value distinction without discriminating frame-independence from recoordinatisation-invariance. The non-degenerate GPS discriminating comparison follows in \S\ref{VFN}.}
Consider a flat, matter-dominated FLRW cosmology, with Ricci scalar $R(t)=\frac{4}{3t^2}$ in cosmic-time gauge. For $t>0$, choose two distinct, locally invertible scalar clocks, $\phi(t)=t$ and $\phi'(t)=t^3$. Expressing $R$ through their inverses gives
\begin{equation}
    R(\phi)=\frac{4}{3\phi^2},\qquad
    R(\phi')=\frac{4}{3\phi'^{\,2/3}}.
\end{equation}
These are two genuinely different functional forms satisfying $R(\phi')=R(\phi)\circ f^{-1}$, where
\begin{equation}
    f:=\phi'\circ\phi^{-1},\qquad f(\phi)=\phi^3
\end{equation}
is the \textit{change of frame-clock map}. The map acts on the clocks' value spaces, not on spacetime; it is the one-dimensional prototype of the inter-frame map $\mathbf{m}$ ( eq. \eqref{eq:m} in \S\ref{VFN}).
At the same physical event labelled by $t=2$, the clocks read $\phi=2$ and $\phi'=8$, but both expressions give $\hat{R}:=R(\phi')|_{\phi'=8}=R(\phi)|_{\phi=2}=1/3$. The functional forms differ, while the scalar's value agrees.

By contrast, a relationally constructed $(0,2)$ tensor $T_{IJ}(\Phi)$ is frame-\textit{dependent}: both its functional form and its component values depend on the frame. The discussion below uses the relational metric $g_{IJ}(\Phi)$ as its main example, but the argument does not depend on this choice.

The next section develops the fibre-bundle construction of these observables, providing the formal setting for the two Views in \S\S\ref{VFN}--\ref{sec3}.

\section{Reference Frames in the Fibre Bundle Formalism}\label{sec1}

In this section, I introduce the fibre-bundle formalism for reference frames and relational observables. Widely used in foundational studies of gauge theories (see e.g. \citealp{healey2007gauging,Weatherall2016-WEAUG}), this formalism has been recently revived in the context of GR by \cite{GomesReprConventions,Gomes25}.

Let $M$ denote the space of models $m$ of the theory.\footnote{Unless otherwise specified, $M=M_{\rm dyn}\subseteq M_{\rm kin}$, where $M_{\rm kin}$ contains kinematically admissible field configurations and $M_{\rm dyn}$ contains solutions of the dynamics \citep{James_Read2023-mk}.} This space can be interpreted as a \textit{fibre bundle} structured by the action of the theory's symmetry group $\mathcal{S}$, generating orbits $\mathcal{O}_m:=\{m_s\mid s\in\mathcal{S}\}$, where $m_s$ is the transform of $m$ by $s$. I denote with $[M]:=M/\mathcal{S}$ its base space, whose elements $[m]$ are equivalence classes containing models related by $\mathcal{S}$-transformations (the action of the symmetry group $\mathcal{S}$).
Within this formalism, selecting a reference frame corresponds to introducing a smooth (local) \textit{section} $\sigma:[M]\to M$, which selects a \textit{single} representative model within each equivalence class. Geometrically, this amounts to specifying a subspace of $M$ that intersects each fibre \( \mathbb{F}_m := \mathrm{pr}^{-1}([m]) \) exactly once, where \( \mathrm{pr} : m\in M \to [m] \in [M] \) is the \textit{projection map}.\footnote{\label{fn:gomes-bridge}The formalism developed in this section is structurally identical to that of \cite{GomesReprConventions}, modulo notational conventions. In \citet[Defs.~3--4]{GomesReprConventions}, $\sigma$ is a \emph{representational convention}; 
Gomes, however, distinguishes between two related objects that I shall deliberately combine into one. He distinguish (i) a \emph{dressing function} $g_\sigma:M\to \mathcal{S}$, taking each model to the group element that gauge-fixes it, and (ii) an \emph{equivalent projection operator} $h_\sigma:M\to M$, $m\mapsto h_\sigma(m):=\sigma([m])$, which sends each model to the unique representative of its orbit on the gauge-fixed surface \citep[Def.~4]{GomesReprConventions}. The relation between the two is $h_\sigma(m)=m^{g_\sigma(m)}$. For the diffeomorphism gauge group, the action of the group on tensorial fields is implemented by pullback. Accordingly, my field-dependent diffeomorphism $f_{\Re_g}$ introduced in \eqref{eqgauge} below plays the role of \emph{both} of Gomes' objects at once: as a diffeomorphism, $f_{\Re_g}\in\mathcal{S}$ is Gomes' dressing function $g_\sigma$; via its pullback action $f^{*}_{\Re_g}$ on the metric, it implements Gomes' projection operator $h_\sigma$, with the relational observable $g_{IJ}(\Re_g)=f^{*}_{\Re_g}(g_{ab})$ being the metric component of the dressed model $h_\sigma(m)$. The compression is harmless for the scope of my paper. Also, the pullback notation is preferred because it is transparent for the GR reader and meshes naturally with \eqref{GI} and the proof of equation \eqref{eq3} below.}  

In gauge-theoretic contexts, the symmetry group \( \mathcal{S} \) generates \textit{gauge orbits} of the bundle: sets of models related by gauge transformations generated by the first-class constraints of the theory. Such $\mathcal{S}-$related models are then physically equivalent: each orbit contains redundant, isomorphic representations of the same physical state selected by the section map $\sigma$—here a \textit{gauge-fixing}. 
This formalism naturally aligns the structure of gauge theories with the relational framework, making the choice of a reference frame formally equivalent to a gauge-fixing \citep{Dittrich2007,Bamonti2023,Bamonti2026}; for applications in cosmology, see \citet{Bamontithebault}.\footnote{See also \citep{Rovelli2014,adlamobservability}.}

For a concrete example, consider \textit{vacuum GR}. Let $M_{\rm dyn}=\mathrm{Lor}(\mathcal{M})$ denote the Lorentzian metrics $\langle\mathcal{M},g_{ab}\rangle$ satisfying the Einstein Field Equations (EFEs) (I will no longer write the manifold $\cal M$ and sometimes suppress the pedix \textit{dyn}, as I will always refer to dynamical models).
The bundle structure arises from the action of the diffeomorphism group \( \mathcal{S} \equiv \text{Diff}(\mathcal{M}) \) and the base space is the set of equivalence classes \([ \text{Lor}(\mathcal{M}) ] := \{ [g_{ab}], g_{ab} \in \text{Lor}(\mathcal{M}) \}\).\footnote{Gribov obstructions generally prevent a global product decomposition $[\mathrm{Lor}(\mathcal{M})]\times\mathrm{Diff}(\mathcal{M})$ \citep{Gribov1978,Henneaux1994-ka}.}
Each equivalence class \([g_{ab}]\) consists of all diffeomorphism-related metrics:
\begin{equation}
[g_{ab}] := \{ g_{ab}, (d^*g)_{ab}, \dots \}, \quad d\in \rm Diff(\mathcal{M}).
\end{equation}

Choose four functionally independent curvature scalars $\Re_g^{(I)}$, $I=1,\ldots,4$, constructed from $g_{ab}$: the \textit{Kretschmann--Komar scalars} \citep{Kretschmann1918,Komar,Bergmann:1960,Bergmann:1962}.\footnote{\cite{Komar} obtains these scalar invariants from the eigenvalue problem \( (R_{abcd}-\lambda(g_{ac}g_{bd}-g_{ad}g_{bc}))V^{cd}=0 \), where \(V^{cd}\) is antisymmetric. The four real invariants determined by this problem are assumed to be functionally independent and identified with $\Re^{(I)}_g$.}  

The set \( \{ \Re^{(I)}_g \} \) defines a \textit{local} diffeomorphism:

\begin{equation}
   \Re_g^{(I)}:= (\Re_g^{(1)}, \cdots \Re_g^{(4)}):U \subseteq \mathcal{M}\rightarrow \mathbb{R}^4, \label{diffeo}
\end{equation}
which assigns a \textit{unique} quadruple of real values to each point in a neighbourhood \( U\subseteq \mathcal{M} \) and supplies a reference frame\textit{ for the metric field itself}.\footnote{\label{jacobian}Functional independence ensures a non-zero Jacobian and local invertibility. It can fail in spacetimes with continuous symmetries, such as Killing symmetries.}
In line with the relational stance, \lq{}\lq{}rather than fixing an observable at specific coordinates, its location is defined relative to features of the state\rq{}\rq{} \citep{Harlow2021AlgebraOD}.

Explicitly, the metric \( g_{ab} \) can be locally expressed in frame-dependent, but gauge-invariant, form using the pullback notation in \eqref{pullback} as:
\begin{equation}
g_{IJ}(\Re_g) := \left[ \Re_g^{-1} \right]^* g_{ab},
\end{equation}
which yields a set of 10 scalar functions indexed by \( I \) and \( J \), constructed from the metric and its derivatives.

Gauge-invariance of $g_{IJ}(\Re_g)$ is ensured by curvature invariants being \textit{dynamically coupled} to the metric \citep{Bamonti2026,BamontiGomes2024}. 
This means that when the pair $(g_{ab},\Re_g^{(I)})$ is a possible solution, then $([d^*g]_{ab},\Re^{(I)})$ is not, for a diffeomorphism $d \in Diff(\mathcal{U})$. Thus, a choice of $\Re_g^{(I)}$ (rather than any of its isomorphic distributions) and initial data give us a \textit{unique} representation for $g_{IJ}(\Re_g)$, ensuring gauge-invariance.
The following holds:
\begin{equation}
 \forall d\in \rm Diff(\mathcal{M}), \quad  [\Re_g^{-1}]^*g_{ab}=[\big(d^*\Re_g\big)^{-1}]^*(d^*g)_{ab}.\label{GI}
\end{equation}

\begin{proof}
\begin{align}
d^*[g_{IJ}(\Re_g)]:=\big[(d^*\Re_g^{(I)})^{-1}\big]^* [d^*g]_{ab}&=\big[(\Re_g^{(I)} \circ d)^{-1}\big]^* [d^*g]_{ab}\notag
\\&= \big[d^{-1} \circ (\Re_g^{(I)})^{-1}\big]^* [d^*g]_{ab}\notag
\\&= \big[(\Re_g^{(I)})^{-1}\big]^* \circ \big[d^{-1}\big]^*[d]^*g_{ab} =: g_{IJ}(\Re_g).
\end{align}
\end{proof}
 
Equation~\eqref{GI} establishes diff-invariance and, for this coupled frame, gauge-invariance.\footnote{In principle, diff-invariance and gauge-invariance are two distinct properties. For dynamically uncoupled frames, independent diffeomorphisms of field or frame may preserve the on-shell conditions and initial data. A relational observable can then be diff-invariant under simultaneous transformations without being gauge-invariant \citep{BamontiGomes2024PARTII,Bamonti2026}. This possibility is excluded for curvature scalar frames, since they are constructed from the metric itself.}

\begin{figure}[h!]
    \centering
    \includegraphics[scale=0.45]{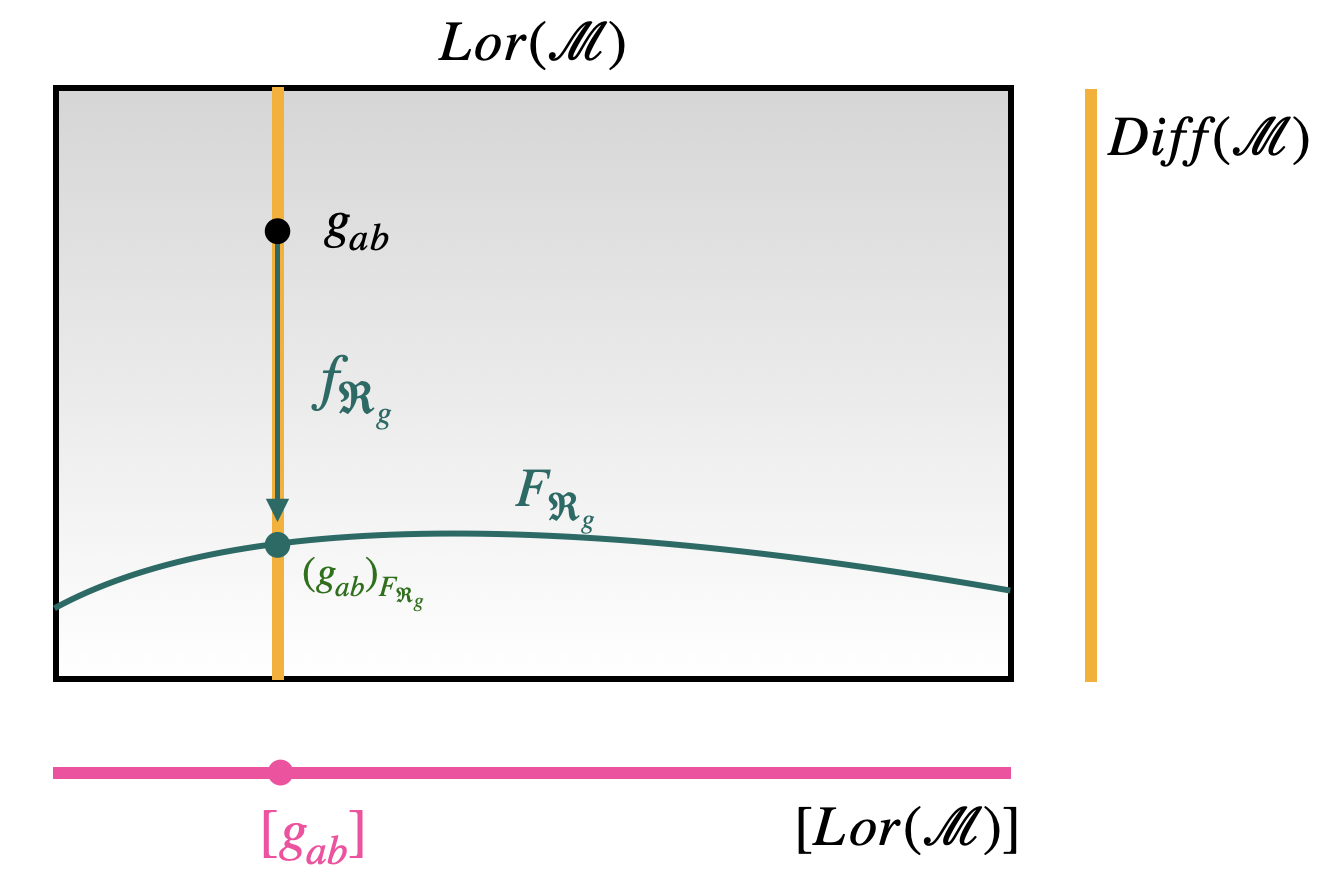}
    \caption{The fibre bundle $\mathrm{Lor}(\mathcal{M})\to[\mathrm{Lor}(\mathcal{M})]$. Each (orange) fibre $\mathbb{F}_g$ is a $\mathrm{Diff}(\mathcal{M})$-orbit of physically equivalent metrics. The frame $\{\Re_g^{(I)}\}$ defines a section whose image is the \lq{}level surface\rq{} $F^{-1}_{\Re_g}(0)$; the pullback by $f_{\Re_g}$ sends each metric to the unique representative $(g_{ab})_{F_{\Re_g}}$ at its fibre's intersection with this image. The base is usually called the \textit{physical state space}.}
    \label{fig1}
\end{figure}

Within the fibre bundle formalism of GR, the choice of \( \{ \Re^{(I)}_g \} \) as a reference frame is expressed as the choice of a section map $\sigma$ that is understood as a gauge choice. 
In practice, this is achieved by choosing a smooth function $F_{\Re_g}:M_{\rm dyn}\to W$, valued in some appropriate space $W$ (typically a space of tensorial densities on $\mathcal{M}$, e.g. $C^{\infty}(\mathcal{M})$ in scalar case), such that:
\begin{equation}\label{eqgauge}  \forall g_{ab} \in Lor, \exists ! \,\rm{( locally)}\, f_{\Re_g}\in \rm Diff(\mathcal{ M})\quad |\quad F_{\Re_g}(f^*_{\Re_g}g_{ab})=0.
\end{equation}
The zero set $F^{-1}_{\Re_g}(0):=\{g\in M_{\rm dyn}\mid F_{\Re_g}(g)=0\}$ is the image of the section. It intersects each fibre once locally, in the absence of stabilisers (cf.\ fn.~\ref{stabilisers}). Namely, it is the submanifold of models satisfying the gauge-fixing condition.\footnote{The choice of a reference frame can be written down in terms of coordinate gauge-fixing \citep{Bamonti2026}. Common examples in GR are harmonic gauge, $F(g)=\partial_\mu(\sqrt{|\det g|}\,g^{\mu\nu})=0$, equivalently $\Box_g x^\nu=0$, and CMC slicing, $F(h^{ij},\pi_{ij})=h_{ij}\pi^{ij}/\sqrt{h}=c(\tau)$, where $\pi^{ij}$ is the ADM momentum density and $h:=\det(h_{ij})$. The latter requires the mean curvature to be spatially constant on each slice $\Sigma_\tau$ of the foliation parametrised by $\tau$, wherever a CMC foliation exists. Both are partial gauge-fixings \citep{landsman2021}.}
 The \textit{field-dependent} diffeomorphism $f_{\Re_g} \in \mathrm{Diff}(\mathcal{M})$ depends on $g_{ab}$ through condition \eqref{eqgauge} and its pullback action \( f_{\Re_g} : g_{ab} \to f_{\Re_g}^* g_{ab} \) projects each model along the gauge orbit onto the image of the section, selecting a unique representative.
Geometrically, the condition \( F_{\Re_g}(f^*_{\Re_g}g_{ab}) = 0 \) defines a \lq{}transversal level surface\rq{} along each gauge orbit: the submanifold where the reference frame condition holds. 
Figure~\ref{fig1} illustrates the construction.\footnote{The fibres are drawn as lines, although gauge orbits need not be one-dimensional.}

\begin{proposition}\label{proposition0}
Given any pair $\big(g_{ab},\Re_g^{(I)}\big)$, $f_{\Re_g}$ yields a frame-dependent representation of the metric $f^*_{\Re_g}g_{ab}$. Under the local identification $U\subseteq \mathcal{M} \to V\subseteq\mathbb{R}^4$ realised by $\{\Re^{(I)}_g\}$, we can write $f^*_{\Re_g}g_{ab}\equiv\big[\Re_g^{-1}\big]^*g_{ab}$. 
This local representation is \textit{unique} and gauge-invariant.
\end{proposition}

\begin{proof}
From the action of the diffeomorphisms $d$,  $f_{d^*\Re_g}=d^{-1}\circ f_{\Re_g}$, we have:
\begin{equation}
f^*_{\Re_g}g_{ab}=f^*_{d^*\Re_g}(d^*g)_{ab}, \forall d \in \rm Diff(\mathcal{M}). \label{eq3} \end{equation}
Now, define
$$
(g_{ab})_{F_{\Re_g}}:=f^*_{\Re_g} g_{ab}.
$$
The previous two equations imply that:
$$
\big((d^*g)_{ab}\big)_{F_{d^*\Re_g}}   =   f^*_{d^*\Re_g}(d^*g)_{ab}=(g_{ab})_{F_{\Re_g}}. \label{eq7}
$$

\end{proof}

Although $f_{\Re_g}$ is model-dependent, its pullback yields the same relational observable throughout each gauge orbit, as shown in \eqref{eq3}.\footnote{Non-trivial stabilisers can make the dressing diffeomorphism $f_{\Re_g}$ non-unique while leaving the dressed field unique. For the curvature frame used here, Killing symmetries obstruct the local invertibility of $\Re_g$. Stabilisers also prevent the full model space from forming a principal $\mathrm{Diff}(\mathcal{M})$-bundle \citep{Komar,Gomes25}. \label{stabilisers}} Yet the gauge-invariant observable \( g_{IJ}(\Re_g) \) remains frame-dependent, since it depends on the chosen section defined by $\Re_g$.\footnote{This also provides a characterisation of what are  called \textit{dressed observables}, see \citet{Harlow2021AlgebraOD,Berghofer2024}.}

We have made explicit the two structures whose ontological interpretation is at issue: frame-free equivalence classes and gauge-invariant, frame-relative observables. The next section situates the established relational framework within physical perspectivalism.

\section{Situating the Debate on Perspectivalism}\label{debunking}

The perspectivalism at issue here is \emph{physical}: perspectives are defined by physical systems serving as reference frames, without requiring conscious observers. This differs from accounts in which perspectives are associated with epistemic communities, conceptual schemes or historically situated scientific practices \citep{Ruyant2020-RUYPRA,Massimi2022-jn,Ruyant2025}. \cite{AdlamPerspect}'s distinction between moderate and strong physical perspectivalism provides the starting point for situating the \emph{View from Nowhere} (\S\ref{VFN}) and the \emph{View from Everywhere} (\S\ref{sec3}).

\emph{Moderate physical perspectivalism} requires empirically meaningful facts to be relativised to physical perspectives, while allowing perspective-neutral facts. A fact is empirically meaningful, in Adlam's terminology, when it describes reality in terms of phenomena that a realistic observer could directly experience; neutral facts typically lack such direct empirical content. \emph{Strong physical perspectivalism} extends the requirement of relativisation to all facts about physical reality, excluding perspective-neutral facts altogether. This distinction concerns facts about physical reality, not the mere availability of mathematical structures in a formalism. In particular, declining to treat a connecting map between perspectives as an additional entity does not remove the physical relations it represents.

This contrast informs the two readings Adlam identifies for  the slogan \lq{}there is no view from nowhere\rq{}. On an empirical reading, a \emph{view} is an empirically meaningful description, necessarily tied to an embodied perspective. The moderate position can therefore endorse the slogan without denying neutral facts. The strong position gives the slogan an ontological reading as well, denying that such facts exist. 
Since the present paper concerns ontological commitments, I adopt the ontological use of \emph{view} when naming the two Views: their disagreement concerns which structures the world contains, while both retain the relational character of empirical access. My use of the term does not contradict Adlam's, it complements it for a different purpose. 
Adlam poses the contrast as follows:
\begin{quote}
[\dots] empirically meaningful descriptions typically have to be relativized to a perspective. But this in and of itself does not tell us whether we should adopt moderate or strong physical perspectivalism. To make that choice, we must decide whether these considerations suggest that all facts about physical reality must be relativized to a perspective, or whether they instead support \textit{the existence of some perspective-neutral facts about physical reality, over and above descriptions relativized to perspectives} [my italics] \citep[p.629]{AdlamPerspect}.
\end{quote}
The meaning of \emph{neutrality} for perspective-neutral facts also needs care. Adlam uses the term for facts not relativised to a perspective; her examples include facts about relations between perspectives and structures such as Minkowski spacetime \citep[p. 640]{AdlamPerspect}. 
Using the refined terminology of \S\ref{terminology}, I distinguish two formal readings of neutrality: \emph{frame-freedom} or \emph{frame-independence}. 

This distinction reveals why frame-freedom defines the contrast between the two Views. Both can admit frame-independent relational structures and could commit ontologically to them. Their disagreement concerns whether a shared frame-free physical situation underlies the relational descriptions. The View from Nowhere admits such a situation alongside the frame-relative ones; the View from Everywhere\textit{ }takes each frame-relative situation to be ontologically comprehensive in its own right. Frame-independence alone does not settle this disagreement.
The relation to physical perspectivalism must therefore be specified. 
When the neutral facts admitted by moderate perspectivalism are represented by frame-free structures, the View from Nowhere accommodates that position. It includes the version Adlam discusses in which perspectival and neutral facts are equally fundamental \citep[p.~632]{AdlamPerspect}, while allowing other assignments of relative fundamentality. The View from Everywhere develops the rejection of neutral ontology on this frame-free reading.

The importance of retaining frame-independence as a separate notion emerges when we ask how perspectives can be connected. Adlam argues that strong perspectivalism cannot supply the required connections, as these necessitate certain neutral facts that serve as a \lq{}perspective-neutral superstructure\rq{} linking the various internal views. Without such structure, facts about any proposed connecting structure would themselves have to be relativised, leading to a further iteration of the problem. Her objection links the ontology of physical perspectives to the conditions of shared knowledge and the formation of epistemic communities. 
Moderate perspectivalism can instead appeal to a neutral structure that explains the relations between perspectives \citep[pp.~635--641]{AdlamPerspect}. The resulting threat of inter-perspectival isolation is the worry relevant here.\footnote{I focus on this structural-connectability objection, asking  whether physically constrained relations between frame-relative observables require a shared frame-free physical situation. Establishing such relations does not by itself explain how agents acquire or share knowledge. Criticisms, such as those related to \textit{epistemic} perspectivalism, are then bracketed due to the narrow scope of the present paper. \label{noepistemic}}
Within the present framework, the question addressed in the next sections is whether the connective role Adlam demands requires \emph{frame-free} structures in the ontology, or whether \emph{frame-independent} relations suffice.

\section{The View From Nowhere}\label{VFN}

The View from Nowhere, as developed here, treats frame-relative situations as aspects of a shared, more comprehensive frame-free physical situation. In the GR construction below, this shared situation is represented by a diffeomorphism-equivalence class of whole models, taken to represent \emph{the} physical situation underlying various possible perspectival descriptions.

For this analysis, I replace the \lq{}Komar\rq{} reference frame $\{\Re_g^{(I)}\}$ of \S\ref{sec1} with the GPS construction of \cite{RovelliGPS}. There is nothing inherently special about either choice. Any well-behaved frame yielding a unique projection $f_{[\bullet]}$ for each isomorphism class is equally suitable. However, GPS frames show how physical devices implement a frame operationally. As with Komar's construction, the resulting observables form a complete set for the corresponding metric--frame situation in the region covered \citep[p.~1182]{Komar}.\footnote{For clarity: \textit{each GPS observable} correspond to a complete set, where the \rq{}set\lq{} refers to the whole family indexed by the infinite values taken by the frame.}

Each GPS frame is represented by four scalar fields, corresponding to the proper-time signals transmitted by four \emph{test} satellites following geodesics of the metric.\footnote{These are reference frames dynamically coupled to the metric without inducing backreaction. They are Coupled Reference Frames (\textbf{CRFs}) in \cite{Bamonti2026,BamontiGomes2024}'s classification.} The satellites start from a common point $O$ and broadcast the proper times elapsed along their geodesics. The signals received at a target point $p$ define a quadruple $\Phi(p):=(\phi^{(1)}(p),\dots,\phi^{(4)}(p))$ that physically coordinatises $p$. Each such quadruple constitutes a local diffeomorphism $\Phi: U\subset\mathcal{M}\to\mathbb{R}^4$, wherever the relevant Jacobian is non-degenerate (cf. fn. \ref{jacobian}).

Now consider two distinct GPS frames: the \emph{red} frame $\Phi_r:=\{\phi^{(I)}_r\}$ and the \emph{blue} frame $\Phi_b:=\{\phi^{(I)}_b\}$. Each provides its own physical parametrisation of the same region $U\subseteq\mathcal{M}$. Still, the two frames are physically distinct: two fleets of satellites.

This setting brings the paper's interpretive question into focus: \emph{how should we interpret distinct relational observables (here $g_{IJ}(\Phi_r)$ and $g_{IJ}(\Phi_b)$) defined relative to different physical frames?} The distinction between the two Views need not be confined to this setting, but becomes especially clear when several material frames are present.

The kinematical space of models of the theory is the Cartesian product
\begin{equation}
M_{\rm kin} \;:=\; \mathrm{Lor}(\mathcal{M}) \,\times\, \mathcal{F}_r \,\times\, \mathcal{F}_b,
\qquad
\mathcal{F}_{r/b} := C^\infty(\mathcal{M},\mathbb{R})^4,
\end{equation}
whose points are ordered triples $m=\langle g_{ab},\phi_r^{(I)},\phi_b^{(I)}\rangle$. The dynamical subspace $M_{\rm dyn}\subseteq M_{\rm kin}$ consists of triples for which $g_{ab}$ satisfies Einstein's equations and each frame satisfies its associated equations of motion (timelike-geodesic propagation of the satellites and null propagation of the radio signals, in the GPS case).

Crucially, the relevant gauge group $\mathrm{Diff}(\mathcal{M})$ acts \emph{diagonally} on the triple. That is:
\begin{equation}
d \cdot \langle g_{ab},\phi_r^{(I)},\phi_b^{(I)} \rangle \;=\; \langle d^{*}g_{ab},\, d^{*}\phi_r^{(I)},\, d^{*}\phi_b^{(I)}\rangle.\label{eq:diagonalaction}
\end{equation}
This diagonality is a consequence of the dynamical coupling of \emph{both} frames to the same metric: independent diffeomorphisms of the three field slots do not generally preserve their solution space \citep{BamontiGomes2024}.

The bundle structure on $M_{\rm dyn}$ is therefore
\begin{equation}
\mathrm{pr}: M_{\rm dyn} \;\longrightarrow\; [M_{\rm dyn}] := M_{\rm dyn}/\mathrm{Diff}(\mathcal{M}),\label{eqbundle}
\end{equation}
whose base $[M_{\rm dyn}]$ is the space of equivalence classes of \emph{whole models}. Each equivalence class
\begin{equation}
[m] \;:=\; \big\{\langle g_{ab},\Phi_r,\Phi_b\rangle,\langle d^{*}g_{ab},d^{*}\Phi_r,d^{*}\Phi_b\rangle,\dots,\,|\, d\in\mathrm{Diff}(\mathcal{M})\big\}
\end{equation}
corresponds to a \emph{single} orbit under the diagonal action \eqref{eq:diagonalaction}. It is essential to emphasise that $[m]$ is \emph{not} the Cartesian product $[g_{ab}]\times[\Phi_r]\times[\Phi_b]$ of three independent equivalence classes (where $[\Phi_{r/b}]$ would denote the Diff-orbit of each frame taken in isolation). Such a product would allow configurations in which the metric and the frames are independently displaced by independent diffeomorphisms, which is precisely what the diagonal action forbids on shell.

\paragraph{Sections and Relational observables.}
Two natural maps on the bundle must be introduced. 

The first is the \emph{restriction} of a model to a sub-system:
\begin{align}
\rho_r &:\, M_{\rm dyn} \to \mathrm{Lor}(\mathcal{M})\times\mathcal{F}_r,
\quad \langle g_{ab},\Phi_r,\Phi_b\rangle \mapsto \langle g_{ab},\Phi_r\rangle,\\[2pt]
\rho_b &:\, M_{\rm dyn} \to \mathrm{Lor}(\mathcal{M})\times\mathcal{F}_b,
\quad \langle g_{ab},\Phi_r,\Phi_b\rangle \mapsto \langle g_{ab},\Phi_b\rangle.
\end{align}
These are simply set-theoretic projections onto a sub-system. They do not break the dynamical coupling between the metric and the two frames in $M_{\rm dyn}$: the dynamics, encoded in the on-shell condition defining $M_{\rm dyn}$, is preserved upstream of the restriction. What $\rho_{r/b}$ does is to look at the part of an on-shell model that involves only the metric and one of the two frames. 
Denote the image of $\rho_{r/b}$ by $M_{\rm dyn}^{(r/b)}\subseteq \mathrm{Lor}(\mathcal{M})\times\mathcal{F}_{r/b}$.

\begin{figure}[h]
    \centering
    \includegraphics[width=0.75\textwidth]{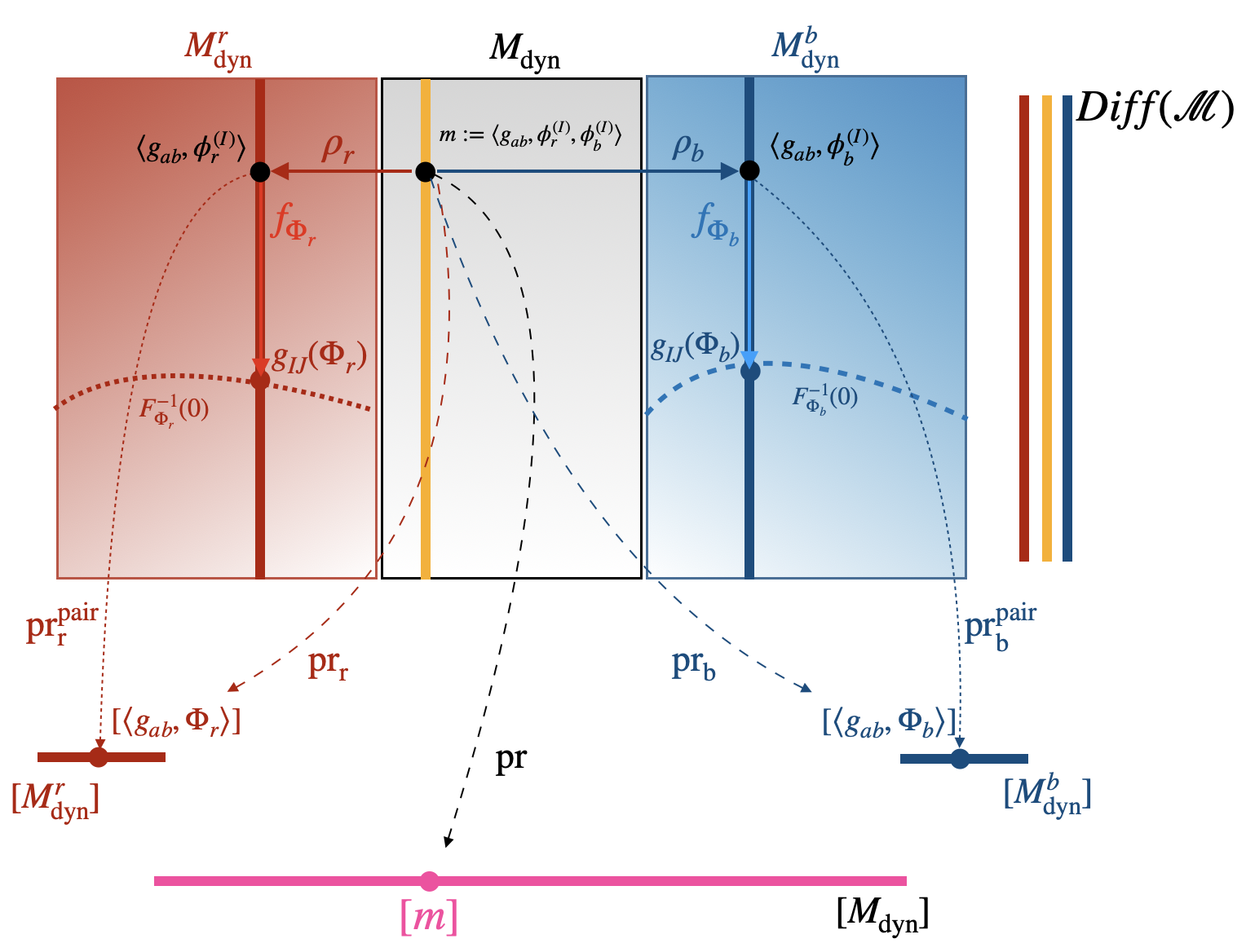}
   \caption{The full-model bundle $M_{\rm dyn}$ and the two pair-model bundles $M_{\rm dyn}^{(r/b)}$. The orange fibre is the diagonal $\mathrm{Diff}(\mathcal{M})$-orbit of the triple; the red and blue fibres are the corresponding diagonal $\mathrm{Diff}(\mathcal{M})$-orbit of the pair-models orbits $\langle g_{ab},\Phi_r\rangle$ and $\langle g_{ab},\Phi_b\rangle$. The restrictions $\rho_{r/b}$ map the full-model orbit onto these pair-model orbits. Within each pair-model fibre, $f_{\Phi_{r/b}}$ selects the representative $g_{IJ}(\Phi_{r/b})$ on $F_{\Phi_{r/b}}^{-1}(0)$. The construction thus reproduces \eqref{eq:extraction}. These observables encode representatives of the pair-model spaces $M_{\rm dyn}^{(r/b)}$, whose classes $[\langle g_{ab},\Phi_{r,b}\rangle] \in [M_{\rm dyn}^{r,b}]$ are obtained through their bundle projections $\mathrm{pr}_{r/b}^{\rm pair}$ that satisfy $\mathrm{pr}_{r/b}=\mathrm{pr}_{r/b}^{\rm pair}\circ\rho_{r/b}$. The full projection $\mathrm{pr}$ instead yields $[m]\in [M_{\rm dyn}]$, whose complete red-frame description requires both $g_{IJ}(\Phi_r)$ and the inter-frame map $\mathbf{m}$ (cf. Proposition~\ref{prop:partial}).}
    \label{fig2}
\end{figure}

The second is the \emph{partial projection} obtained by composing the restriction with the quotient by the diagonal Diff-action on the corresponding pair:
\begin{equation}
\mathrm{pr}_{r/b}:\, M_{\rm dyn} \to \big[M_{\rm dyn}^{(r/b)}\big] := M_{\rm dyn}^{(r/b)}/\mathrm{Diff}(\mathcal{M}),
\quad \langle g_{ab},\Phi_r,\Phi_b\rangle \mapsto [\langle g_{ab},\Phi_{r/b}\rangle].\label{eq:partialproj}
\end{equation}
Here $[\langle g,\Phi_{r/b}\rangle]$ is the equivalence class of the dynamically coupled metric--frame \textit{pair} under the diagonal Diff-action on that pair. The pair quotients $[M_{\rm dyn}^{(r/b)}]$ differ from the full-model quotient $[M_{\rm dyn}]$: the partial projection $\mathrm{pr}_r$ identifies on-shell triples with the same $\langle g_{ab},\Phi_r\rangle$ but differ only in the configuration of $\Phi_b$, and analogously for $\mathrm{pr}_b$. 
Equivalently, we can work directly with the two pair-model bundles $M_{\rm dyn}^{(r/b)}\to[M_{\rm dyn}^{(r/b)}]$. Their bundle projections $\mathrm{pr}_{r/b}^{\rm pair}$ satisfy $\mathrm{pr}_{r/b}=\mathrm{pr}_{r/b}^{\rm pair}\circ\rho_{r/b}$. The restrictions $\rho_{r/b}$ are equivariant maps from triples to pairs; they do not select representatives within the full-model fibres. Both descriptions are equivalent for the present purposes (see Fig. \eqref{fig2}).

A \textit{choice} of a reference frame corresponds, as in \S\ref{sec1}, to a (local) section of the bundle picking out a frame to play the role of reference. For the construction of $g_{IJ}(\Phi_r)$, the relevant local section belongs to the pair-model bundle $M_{\rm dyn}^{(r)}\to[M_{\rm dyn}^{(r)}]$.
Selecting the red frame means means first extracting the pair
$(g_{ab},\Phi_r)\in M_{\rm dyn}^{(r)}$ via $\rho_r$, and then performing the gauge-fixing analogous to \eqref{eqgauge} via associated field-dependent diffeomorphism $f_{\Phi_r}$ projecting each fibre onto the image of the section, namely $F^{-1}_{\Phi_r}(0)$. 
Concretely, the field-dependent diffeomorphism $f_{\Phi_r}$ acts on the restricted model $\langle g_{ab},\Phi_r \rangle$ and yields the gauge-invariant relational observable (cf. Prop. \eqref{proposition0})
\begin{equation}
g_{IJ}(\Phi_r) \;:=\; f^{*}_{\Phi_r}\,g_{ab} \;\equiv\; [\Phi_r^{-1}]^{*}\,g_{ab}.\label{eq:redobs}
\end{equation}

Schematically, the extraction of the observable from the full model thus proceeds as follows:
\begin{equation}
m=\langle g_{ab},\Phi_r,\Phi_b\rangle \;\xrightarrow{\;\rho_r\;}\; \langle g_{ab},\Phi_r\rangle \;\xrightarrow{\;f_{\Phi_r}\;}\; g_{IJ}(\Phi_r).\label{eq:extraction}
\end{equation}
The same construction with the blue frame yields $g_{IJ}(\Phi_b):=[\Phi_b^{-1}]^{*}g_{ab}$, factoring through $\rho_b$.

\paragraph{Relational Observables and Equivalence Classes.}

Because $\rho_r$ omits the blue frame before $f_{\Phi_r}$ acts, $g_{IJ}(\Phi_r)$ depends only on the metric and the red frame. This has a direct consequence for what the observable determines.

Two on-shell triples $\langle g_{ab},\Phi_r,\Phi_b\rangle$ and $\langle g_{ab},\Phi_r,\Phi_b'\rangle$, with the same metric and the same red frame but different blue frames, yield the \emph{same} value for $g_{IJ}(\Phi_r)$. 
Yet, by the diagonality of the gauge action \eqref{eq:diagonalaction}, these two triples generally belong to \emph{distinct} equivalence classes $[m]\neq [m']$.\footnote{Provided the change $\Phi_b\to\Phi_b'$ is not implemented by a diffeomorphism that simultaneously fixes both $g$ and $\Phi_r$, which would require $d$ to belong to the \textit{stabiliser} of $(g,\Phi_r)$. For generic configurations this stabiliser is trivial.} 
Crucially, as I will examine in more details in \S\ref{sec:VFN.empirical}  below, this shows that the whole-model class $[m]$ carries strictly more \lq{}physical information\rq{} than either of its two partial equivalence classes $[\langle g_{ab},\Phi_{r/b}\rangle]$.\footnote{\label{fninfo}Throughout, `physical information' means gauge-invariant content, without reference to Shannon information, algorithmic complexity or epistemic states. To say that $[m]$ carries more physical information than $g_{IJ}(\Phi_r)$ is to say that $[m]$ encodes gauge-invariant content that $g_{IJ}(\Phi_r)$ alone lacks---as established by Proposition~\ref{prop:partial} below.}
Each relational observable captures only the gauge-invariant content of the corresponding partial projection \eqref{eq:partialproj}---a sub-part of the full content of $[m]$. We can summarise this as follows.

\begin{proposition}\label{prop:partial}
The relational observable $g_{IJ}(\Phi_r)$ provides (locally, in the region where $\Phi_r$ is a valid reference frame) a unique gauge-invariant representation of the equivalence class $\mathrm{pr}_r(m) = [\langle g_{ab},\Phi_r\rangle] \in [M_{\rm dyn}^{(r)}]$ of the \emph{sub-model} consisting of the metric and the red frame alone. It does \emph{not} uniquely represent the whole equivalence class $[m]=[\langle g_{ab},\Phi_r,\Phi_b\rangle]\in [M_{\rm dyn}]$. The same holds, \textit{mutatis mutandis}, for $g_{IJ}(\Phi_b)$.\footnote{With only one frame coupled to the metric, the partial and full projections coincide: $g_{IJ}(\Phi)$ captures $[\langle g_{ab},\Phi\rangle]$. This case conceal interesting features, such as the information gap discussed in \S\ref{sec:VFN.empirical}.}
\end{proposition}

 The fibre $[m]\in [M_{\rm dyn}]$ is not uniquely picked out by either observable alone. In the language of Komar's classical formulation, $g_{IJ}(\Phi_r)$ is a complete set of observables \emph{only} for the partial-projection class $[\langle g_{ab},\Phi_r\rangle]$, not for the whole-model class $[m]$.

To recover the full gauge-invariant content of $[m]$ in a red-frame description, one must supplement $g_{IJ}(\Phi_r)$ with the local diffeomorphism between the value spaces of the two frames,
\begin{equation}
\mathbf{m}\;:=\;\Phi_b\circ\Phi_r^{-1}\;:\; \Phi_r(U)\subseteq\mathbb{R}^4 \to \Phi_b(U)\subseteq\mathbb{R}^4,\label{eq:m}
\end{equation}
which encodes precisely how the blue frame labels the points already labelled by the red frame.\footnote{The map $\mathbf{m}$ is what \cite{BamontiGomes2024PARTII} term an \emph{external diffeomorphism}. It is well defined wherever the relevant Jacobians are non-degenerate. I shall return to its full interpretive role in \S\ref{sec3}. For present purposes, it functions as the missing piece of gauge-invariant data needed to reconstruct $[m]$ from a single relational observable.} 
The map \textbf{m} functions \textit{analogously} to a passive transformation, but with an important formal difference: unlike an \textit{ordinary} passive diffeomorphism, \textbf{m} is not a pure redescription of the same tensorial fields in different coordinates, but it produces a genuine change of physical systems, whose dynamics constrain the admissible map from a relational observable to another according to: 
\begin{equation}
g_{IJ}(\Phi_r)=\textbf{m}^*g_{IJ}(\Phi_b).
\end{equation}
This is verified directly:
\begin{align}
(\textbf{m})^*[g_{IJ}(\Phi_b)]&=(\textbf{m})^*\big[ (\Phi_b^{-1})^*g_{ab}\big] \notag\\ 
&= \big[ \Phi_b^{-1}\circ \textbf{m}\big]^*g_{ab} \notag\\ 
&= \big[\Phi_b^{-1 }\circ \Phi_b\circ \Phi_r^{-1}\big]^*g_{ab} \notag \\
&=\big [ \Phi_r^{-1}\big]^*g_{ab} \notag \\
&= g_{IJ}(\Phi_r).
\end{align}

With $\mathbf{m}$ in hand, the pair $\big(g_{IJ}(\Phi_r),\,\mathbf{m}\big)$ \emph{does} (at least locally) uniquely identify $[m]$, and likewise for $\big(g_{IJ}(\Phi_b),\,\mathbf{m}^{-1}\big)$.

In compact form, the structure can therefore be summarised as follows:
\begin{equation}
\underbrace{[m]}_{\text{whole-model class}}\;\longleftrightarrow\;
\big(g_{IJ}(\Phi_r),\,\mathbf{m}\big)\;\longleftrightarrow\;
\big(g_{IJ}(\Phi_b),\,\mathbf{m}^{-1}\big),
\end{equation}
while
\begin{equation}
\underbrace{[\langle g_{ab},\phi_r\rangle]}_{\text{sub-model class}}\;\longleftrightarrow\;g_{IJ}(\Phi_r),
\qquad
\underbrace{[\langle g_{ab},\phi_b\rangle]}_{\text{sub-model class}}\;\longleftrightarrow\;g_{IJ}(\Phi_b).
\end{equation}

Figure~\ref{fig2} schematises the bundle. 

\subsection{The Ontological Thesis}\label{sec:VFN.thesis}

With the formal structure in place, the interpretive question of this section can now be sharply posed: \emph{which structures in this hierarchy do we associate with an ontological commitment?}
A minimal necessary requirement is that any structure to which one commits ontologically must be gauge-invariant.  
A clarification is in order. In speaking of \lq{}committing ontologically to\rq{} relational observables, equivalence classes and the like, I mean treating these mathematical structures as formal models of real physical situations. The commitment concerns the situations represented.

The hierarchy of gauge-invariant structures of \S\ref{VFN} offers the following natural candidates:
\begin{itemize}
    \item the relational observables $g_{IJ}(\Phi_r),g_{IJ}(\Phi_b)$.\footnote{By Proposition~\ref{prop:partial}, each $g_{IJ}(\Phi_{r/b})$ is in canonical bijection with the corresponding partial class $[\langle g,\Phi_{r/b}\rangle]$. Here, I decide to use the observables rather than the corresponding partial classes as candidates. This reflects the relevant disagreement between the two Views concerning commitment to the frame-free class $[m]$ and the relational observables.}
    \item the equivalence class $[m]$ of the whole model.
\end{itemize}

The inter-frame map $\mathbf{m}$ is another gauge-invariant candidate.\footnote{Under the diagonal $\mathrm{Diff}(\mathcal{M})$-action, $\mathbf{m}^d=(\Phi_b\circ d)\circ(\Phi_r\circ d)^{-1}=\Phi_b\circ\Phi_r^{-1}=\mathbf{m}$.} It represents a \textit{physically constrained} relation between the observables: the relational fact that the two observables are connectable via a physical map.  Both Views can accept this candidate; their disagreement concerns commitment to \textit{frame-free} class structures alongside the relational observables. So, for the present purpose, I will focus only on commitments on relational observables and equivalence classes, leaving open the commitment on \textbf{m}.\footnote{\label{statusm}What remains open is whether relations are additional entities, not whether the frames are physically related. The implications for inter-perspectival connectability are examined in \S\ref{sec3}.}

The thesis characterising the View from Nowhere can be stated as follows:

\begin{description}
\item[\textbf{The View from Nowhere:}] \emph{The most comprehensive physical situation is represented by the frame-free equivalence class $[m]$ of whole models $\langle g,\Phi_r,\Phi_b\rangle$. Relational observables $g_{IJ}(\Phi_r),g_{IJ}(\Phi_b),\dots$ are gauge-invariant perspectival `points of view' on $[m]$, each \lq{}Komar-completely\rq{} representing only an ontologically partial physical situation corresponding to a partial sub-model class but never, individually, $[m]$ itself.}
\end{description}

Under the frame-free reading of perspective-neutrality, the View from Nowhere admits the equal-fundamentality option discussed by \citet[p.~632]{AdlamPerspect}: perspectival and frame-free facts can be equally fundamental. Identifying $[m]$ with a shared, more comprehensive physical situation is the specific GR interpretation developed here.
An even more deflationary variant provides a GR analogue of what \citet[\S 4]{AdlamPerspect} calls a \emph{Cartesian--Hegelian} position, on which frame-free structures are the \emph{only} fundamental ones and perspectival quantities are reducible to them. My aim here is not to adjudicate between these variants but to address the broader family that retains $[m]$ in its ontology, to which both belong. 

Two clarifications are needed: one concerns the notions used to state the thesis, the other the status of the frame-free situation at stake. 

\paragraph{Gauge-invariant content, completeness, comprehensiveness.}
The thesis just stated leans on three notions that must be kept apart. 

\emph{Gauge-invariant content (or information)} is a property of a structure of the bundle: by Proposition~\ref{prop:partial}, $[m]$ carries strictly more of it than any single relational observable $g_{IJ}(\Phi)$. This is a formal fact, prior to and independent of any interpretive choice.

\emph{Komar-completeness} has to be understood not as a one-place property but as a two-place relation between an observable (or set of observables) and a \emph{target} situation: a set is Komar-complete \emph{for} $Y$ when it `uniquely determines or characterises' $Y$, capturing its whole gauge-invariant content. The bare question \lq{}is $g_{IJ}(\Phi)$ Komar-complete?\rq{} is thus ill-formed until the target is fixed---$g_{IJ}(\Phi)$ is Komar-complete \emph{for} the partial-projection class $[\langle g,\Phi\rangle]$, but not \emph{for} $[m]$ (Proposition~\ref{prop:partial}).

\emph{Comprehensiveness} concerns the ontological status of a \textit{physical situation}: a situation is comprehensive when it is not taken to be a partial aspect of any further physical situation admitted by the ontology. I will use 'ontologically partial' to mean non-comprehensive, and distinguish it from 'partial' simpliciter which means that a structure captures only part of the gauge-invariant content of a specified target. Hence, comprehensiveness implies neither maximal gauge-invariant content nor greater fundamentality: a fundamental situation may still be an aspect of a more comprehensive one.

This is the sense in which the thesis above is an \emph{ontological} reading and not a consequence of the formalism alone: the View from Nowhere takes the further step of treating $[m]$ as a genuine situation---indeed the most comprehensive one---of which each $g_{IJ}(\Phi)$ is an ontologically partial, frame-relative manifestation. On that reading $[m]$ \emph{is} understood as more comprehensive, but in virtue of the ontological commitment, not in virtue of the content gap of Proposition \ref{prop:partial}, which on its own is silent on the matter. \S\ref{sec3} will decline precisely this step.

\paragraph{The physical situation is $[m]$, not $[g_{ab}]$.} Seduced by the analogy with the approach to coordinate-based descriptions in physics, the following is often a tempting reading of the formalism. Since every solution of the EFEs is fully characterised by its metric content, one might be tempted to read $g_{IJ}(\Phi_r)$ and $g_{IJ}(\Phi_b)$ as two `physical parametrisations' of the same metric solution, representing the full physical situation. 
On this reading, the shared physical situation would be $[g_{ab}]$, the diffeomorphism-equivalence class of the metric alone, with the two frames playing the role of mere operational tools used to coordinatise it. 
This reading captures something true: the two relational observables \textit{do} encode the \lq{}same\rq{} (up to diffeomorphisms) solution $g_{ab}$  of the EFEs described through two different frames. However, associating to this same metric structure the shared physical situation is misleading, for two convergent reasons.

Firstly, once the frames are included in the model space $M_{\rm dyn}$, the gauge group $\mathrm{Diff}(\mathcal{M})$ acts diagonally on the \textit{coupled} triple $m=\langle g,\Phi_r,\Phi_b\rangle$, as in \eqref{eq:diagonalaction}. The equivalence class $[m]$ captures the full gauge-invariant content of the model, whereas $[g_{ab}]$ captures only its metric content.\footnote{\label{gravframes}Crucially, this precise argument is \textit{invisible} in the setting of \S\ref{sec1}. In that case, the metric is the only field in the model and the frames are constructed from it. The shared physical situation is therefore represented by $[g_{ab}]$, regardless of how many such frames are used.} 
Given a triple $\langle g,\Phi_r,\Phi_b\rangle$, the triple $\langle d^{*}g,\Phi_r,\Phi_b\rangle$, sharing $[g_{ab}]$ but related by a non-diagonal transformation, generally is \emph{not} an on-shell solution of the same theory. The same point can also be made within the on-shell space: two different on-shell triples can also share $[g_{ab}]$ \emph{without} going off-shell. In this case, however the difference between the two triples must lie in the \emph{initial conditions} of the frames.  Keep $g_{ab}$ and $\Phi_r$ fixed and vary the blue frame's admissible initial data \citep[\S~1]{RovelliGPS}, obtaining another on-shell model $\langle g,\Phi_r,\Phi_b'\rangle$  with $\Phi_b'\neq\Phi_b$ on the common domain $U$. The two on-shell models $\langle g,\Phi_r,\Phi_b\rangle$ and $\langle g,\Phi_r,\Phi_b'\rangle$ share $[g_{ab}]$, but belong to genuinely distinct equivalence classes $[m]\neq[m']$.\footnote{A diagonal diffeomorphism $d:U\to U$ relating the two triples would have to satisfy $d^*\Phi_r=\Phi_r$. Since $\Phi_r$ is invertible on $U$, this requires $d=\mathrm{id}_U$, which cannot map $\Phi_b$ to $\Phi_b'\neq \Phi_b$. Thus the two configurations belong to distinct gauge orbits in the local construction, even though their metrics coincide.} The difference lies in physical frame data that the metric class alone does not capture.

\emph{Secondly, frames are physical structures, not coordinates}. Although test frames do not backreact on the metric, they remain physical components of the model: they have initial conditions, \textit{dynamical equations} to be considered in addition to the EFEs, and a physically constrained translation map encoded in $\mathbf{m}$.  Identifying the complete physical situation with $[g_{ab}]$ alone would omit these physical frame configurations.
On the View from Nowhere considered here, the complete shared physical situation includes these frame configurations and is therefore represented by the full-model class $[m]$. 
The metric class $[g_{ab}]$ retains a derivative status in this picture: it is the gauge-invariant projection of $[m]$ onto its metric component. It retains the \textit{geometry}, but it is not, by itself, the physical situation that grounds the two relational observables.  The same reasoning applies to the partial gauge-invariant classes $[\langle g,\Phi_r\rangle]$ and $[\langle g,\Phi_b\rangle]$. Each captures only one metric-frame sub-system. Treating either as the physical situation of the \emph{whole model} would relegate the omitted frame to \lq{}surplus formal structure\rq{}, contrary to its status as a dynamically coupled physical component.  
Furthermore, when we consider two on-shell triples with coupled frames, by Proposition~\ref{prop:partial}, distinct $[m]\neq[m']$ may project to the same $[\langle g,\Phi_r\rangle]$, differing only in $\Phi_b$. Only $[m]$ contains the \textit{full} gauge-invariant content of the model.

\subsection{Empirical Access and Ontological Commitment}\label{sec:VFN.empirical}

Advocates of the View from Nowhere sustain that $[m]$, as a frame-free structure, is not directly accessible to measurement. Empirical access to physics is always frame-explicit: an experimenter measures values of relational observables $g_{IJ}(\Phi)$ for some operationally instantiated frame, never the equivalence class $[m]$ as such.\footnote{The relational character of empirical data has been exhaustively documented in the literature. See e.g.\ \citet[p.~128]{Anderson1967-en}, \citet[p.~298]{Rovelli1991}, \citet[p.~1]{Landau1987-fh}, and \citet{HOLE}'s \emph{immanent} conception of empirical distinctness. The point goes back to Einstein's point-coincidence argument \citep{Einstein1916,Einstein1919,Giovanelli2021,BamontiGomes2024}.} The empirically direct inaccessibility of $[m]$ is therefore not a contingent shortcoming of the View from Nowhere, but a defining feature of it. 

Two remarks on what this implies are in order.

\paragraph{Partial Information and Reconstruction.} The formalism of \S\ref{VFN} clarifies the moderate perspectivalist position that the empirical content of $[m]$ could actually be (indirectly) recovered from access to relational observables, given their inter-frame maps \citep[pp.637-39]{AdlamPerspect}. 

Proposition~\ref{prop:partial} reveals that there is, in a precise formal sense, \emph{more physical information} in $[m]$ than is captured by any individual relational observable. Nonetheless, the joint structure $(g_{IJ}(\Phi_r),\mathbf m)$ fully encodes the gauge-invariant content of $[m]$.\footnote{The formal reconstruction does not itself establish the empirical availability of those data.} Hence, reconstructing this content through relational measurements counts as \textit{indirect} empirical access to the physical situation represented by $[m]$.\footnote{\label{fn:VFN-frame-indep}Within the View from Nowhere, the coincidence of values across frames does not make a frame-independent observable a more encompassing representation of $[m]$: when constructed from one metric--frame pair, it still lacks the other frame's data. The most comprehensive physical situation remains the one represented by $[m]$; frame-independence is a property an observable may have, not a route to a deeper physical situation. For the analogous point in the View from Everywhere, see fn.~\ref{fn:VFE-frame-indep}.}

However, this reconstruction proceeds through frame-explicit structures, while the View from Nowhere also commits to $[m]$ as a frame-\textit{free} entity. The question is therefore what additional work, if any, this commitment performs beyond that already performed by the relational observables and their inter-frame map.

\paragraph{Cheapness.} Committing ontologically to $[m]$ could be dismissed (for instance, by a proponent of the View from Everywhere) as \emph{metaphysically cheap} insofar as its empirical role is exhausted by relational observables and their interconnection maps. Calling the reconstruction of its content \lq{}indirect access\rq{} does not, by itself, justify this further commitment.\footnote{The objection concerns the addition of a frame-free entity without additional empirical content, not indirect access as such. This use of \lq{}cheap\rq{} adapts the perspicuity objection to \textit{external} sophistication discussed in \cite{Martens2020}, reframing it here to address empirical redundancy.\label{fn:cheapness}}
The position bears a qualified structural analogy to Newton's commitment to the reality of absolute space, despite its inaccessibility to direct observation \citep{Brown2013}.
Newton appealed to effects of rotation, such as the concavity of the water in the rotating-bucket argument, to infer true rotation relative to absolute space \citep{newton1999principia}. The analogy concerns this indirect route to a structure that is not itself directly observed: the View from Nowhere similarly treats relational measurements as access to an underlying frame-free situation. Whether or not one judges this analogy fair, it captures the methodological tension at issue: whether the frame-free posit contributes any explanatory role \textit{beyond} that supplied by relational observables and their inter-frame maps. 

The View from Everywhere (\S\ref{sec3}) will challenge whether connecting perspectives requires such a posit.

\section{The View From Everywhere}\label{sec3}

The View from Everywhere builds upon the formalism developed in \S\ref{VFN}: the bundle $M_{\rm dyn}$, the diagonal Diff-action, the projections, the relational metric observables obtained from sections of the pair-model bundles, the inter-frame map $\mathbf{m}$ are all in place. What changes is which of these structures bears ontological weight.
Where the View from Nowhere commits to $[m]$ as a shared frame-free physical situation, the View from Everywhere treats it as a formal structure of the bundle.

The reconstruction established in \S\ref{sec:VFN.empirical} remains available: $(g_{IJ}(\Phi),\mathbf m)$ still encodes the full gauge-invariant content of $[m]$. Interpreting this reconstruction as indirect access to a further physical situation, however, requires the ontological commitment the View from Everywhere rejects. It therefore recognises no frame-free physical referent to be accessed, directly or indirectly.
On this reading, the physical situations the theory countenances are exhausted by the relational observables themselves---each gauge-invariant, predictively Komar-complete for the corresponding metric-frame situation, and operationally instantiated by an actual frame.\footnote{Empirically, if a measurement of $g_{11}(\Phi_r)$ yields $1.732$ and a measurement of $g_{11}(\Phi_b)$ yields $2.041$, two physical measurements have been performed, with two different physical apparatuses, yielding two different gauge-invariant numerical results.}  

\begin{description}
\item[\textbf{The View from Everywhere:}] \emph{Each relational observable $g_{IJ}(\Phi_r), g_{IJ}(\Phi_b), \dots$ constitutes a Komar-complete description of a comprehensive and fundamental physical situation that is not ontologically partial with respect to any more comprehensive one. Relational observables define distinct relational realities, and each individual observable exhausts its physical situation. The frame-free equivalence class $[m]$ retains only formal status: it is a quotient construction of the bundle, not a physical entity.}
\end{description} 

On the View from Everywhere, each relational observable characterises a physical situation individuated relative to its frame. Such situations are also admitted by the View from Nowhere; the disagreement concerns whether they are aspects of a further shared frame-free situation. The warning in \S\ref{VFN} concerned treating a partial class as the situation of the whole model. The View from Everywhere instead retains $[m]$ as a formal class without admitting that additional physical referent.

Both Views accept that $g_{IJ}(\Phi_r)$ is Komar-complete for $[\langle g,\Phi_r\rangle]$, but not for $[m]$ (Proposition~\ref{prop:partial}). Their disagreement concerns ontological comprehensiveness, in the sense fixed in \S\ref{sec:VFN.thesis}. The View from Nowhere treats the metric--frame situation as \emph{non-comprehensive}: it is an aspect of the shared physical situation represented by $[m]$. The View from Everywhere takes it to be comprehensive in its own right, while retaining the difference in gauge-invariant content.

\subsection{Towards a purely relational reality: Relality}\label{sec:VFE.relality}

The View from Everywhere takes \textit{each} metric--frame situation characterised by $g_{IJ}(\Phi)$ to be ontologically comprehensive. 
Recognising this status for the red-frame situation does not preclude recognising it for the blue-frame situation or retaining the physical relations between them.

We consider, as a derived structure, the totality of relational metric observables equipped with the inter-frame maps. This structure was already encountered in \S\ref{VFN} as the joint structured collection $(g_{IJ}(\Phi_r),\mathbf{m})$ that fully encodes the gauge-invariant content of $[m]$. It may be explicitly written out using set notation as:
\begin{equation}
\{g_{IJ}\} \;:=\; \big\{ g_{IJ}(\Phi_r), g_{IJ}(\Phi_b),\dots \big\}.
\label{eq:gIJclass}
\end{equation}  
The collection of metric observables alone need not determine its inter-frame maps, so they have to be understood as already included in  \eqref{eq:gIJclass}.\footnote{\label{fn:interframe-data} Write $g_r:=g_{IJ}(\Phi_r)$ and $g_b:=g_{IJ}(\Phi_b)$. If $\mathbf m^*g_b=g_r$ and $h$ is a non-trivial isometry of $g_b$, then $(h\circ\mathbf m)^*g_b=g_r$ as well. The same metric observables are therefore compatible with different inter-frame maps.} 
 
 Even though encoding the same gauge-invariant content,  the set $\{g_{IJ}\}$ differs from the equivalence class $[m]$ in its construction.  
 The class $[m]$ consists of gauge-\emph{variant} models related by a physical equivalence relation (the diagonal active diffeomorphism).  
 The set $\{g_{IJ}\}$ consists of gauge-\emph{invariant} observables and their inter-relations under the action of $\mathbf{m}$, which is \textit{not} a physical equivalence relation: $\mathbf{m}$ relates the observables without identifying them as representations of a single physical situation.\footnote{In \citet[p.~954]{Belot2017}'s terminology, $\mathbf{m}$ is a \emph{physical symmetry}---an isomorphism linking solutions that represent distinct \emph{possibilia}, as opposed to a \emph{gauge symmetry}, which relates representations of the same physical situation.} 

The View from Everywhere treats both $\{g_{IJ}\}$ and $[m]$ as formal structures, without taking either to represent a further physical situation.\footnote{The structured set $\{g_{IJ}\}$ \emph{supervenes} on its constituent metric observables and their specified interrelations in a strongly reductive sense. On this interpretation, higher-level (supervenient) properties are not autonomous entities but supervene from systematic patterns within the underlying base-level entities. A more neutral stance on the ontological status of supervenient entities is also available---see \citet{Butterfield2011, Dewar2019}. For an overview, see \citet{sep-supervenience}.} Ontological commitment is to the individual frame-relative situations; any additional commitment to $\mathbf m$ concerns their interrelations, without making any of them ontologically partial (cf. fn.~\ref{statusm}).\footnote{\label{fn:VFE-frame-indep} Neither agreement in values across frames nor the frame-independence of inter-frame maps entails a more comprehensive physical situation (see \S\ref{terminology}).} I call the structured collection \textit{Relality}, a portmanteau of \textit{relational} and \textit{reality}.

Its informational equivalence with $[m]$ ensures that both Views accept that the same relational data determine the gauge-invariant content of the theory; only the View from Nowhere takes that content to belong to a shared, more comprehensive physical situation. The gauge-invariant content of $\{g_{IJ}\}$ is thus the joint content of distinct situations and their interrelations. What the View from Nowhere interprets as a stand-alone frame-free entity, the View from Everywhere presents through a relational plurality.

\subsection{Inter-perspectival Connectability and Perspective-Neutrality}

The inter-frame map addresses the ontological question isolated in \S\ref{debunking}: whether physically connected frame-relative situations require a shared frame-free situation. Here, \emph{inter-perspectival connectability} means translation between relational observables through maps constrained by the configurations and dynamics of physical frames. Acquiring and exchanging measurement records requires further conditions not analysed here (cf. \eqref{noepistemic}).\footnote{Adlam's \lq{}first-person-plural\rq{} view associates perspectives with epistemic communities: knowledge can extend beyond that of individual members while remaining community-relative \citep[pp.~639--641]{AdlamPerspect}. Her objection to its viability within strong physical perspectivalism concerns the physical interactions and information-sharing relations needed to form such a community. Relality specifies connected physical situations, without modelling the formation of an epistemic community. Neither the existence of its inter-frame maps nor the refusal to treat the collection as an additional entity establishes or excludes a community-relative epistemic interpretation.}

On the common domain where both frames are locally invertible (cf. fn.~\ref{jacobian}), given the red-frame metric observable and the inter-frame map $\mathbf{m}$, the blue-frame observable can be reconstructed. Under the View from Everywhere, the observables so connected characterise distinct, comprehensive physical situations. This is physical covariance under changes of physical reference frame, with gauge-invariance preserved.\footnote{\citet{Wallace2024} discusses the objection that fixing a reference frame reintroduces gauge-dependence under the guise of frame-dependence (cf. \citealp{thiemann26}). This objection fails to recognise the crucial difference between frame-dependence and gauge-dependence. Frame choice indeed can make physics frame-dependent, but the resulting observables remain gauge-invariant; the distinction is already apparent in \citet[pp.~1186--1187]{Komar}. Equally, gauge-invariance does not establish frame-freedom or frame-independence. Adlam's claim that facts about complete observables are \lq{}\lq{}perspective neutral in the strongest possible sense\rq{}\rq{} \citep[p.~630]{AdlamPerspect} elides this distinction. A relational construction does not, by itself, settle whether the represented facts are perspective-neutral in the intended sense.}

The map $\mathbf m=\Phi_b\circ\Phi_r^{-1}$ is frame-\textit{independent} in the sense specified in \S\ref{terminology}: with the blue fields held fixed, changing the dressing frame preserves their scalar values at corresponding events. In fact, the maps satisfy $\mathbf m_{b|r}=\mathbf m_{b|y}\circ\mathbf m_{y|r}$ for any admissible intermediate \lq{}yellow\rq{} frame $\Phi_y$. Componentwise, this expresses precisely the scalar value-invariance. No privileged dressing frame is required. The map is nevertheless not frame-free, since its definition explicitly involves the frames it relates.

Importantly, whether $\mathbf m$ represents an additional entity is a separate question. One may choose to commit ontologically to $\mathbf{m}$: this yields an ontology equipped with a \emph{frame-independent} (though not frame-free) connective entity. Or one may decline such ontological commitment to \lq{}relations\rq{} and treat $\mathbf{m}$ as a \textit{physically constrained} structural feature of the formalism: in that case, $\mathbf{m}$ continues to perform its connective work without being  added to the ontological inventory. 

On the frame-free reading of perspective-neutrality, this discussion provides a constructive counterexample to the claim that structural connections between perspectives require neutral structures in the ontology. Adlam's objection concerns neutral facts about connections: facts that hold without being relativised to a perspective. Declining to treat a relation \textit{itself} as an entity does not remove the fact that the connection exists \citep[pp.~635--636]{AdlamPerspect}. In either case, the inter-frame map $\mathbf{m}$ supplies physically constrained connections between relational observables, and without commitment to a shared frame-free physical situation, thus vindicating strong perspectivalism under that qualified reading.\footnote{An analogy with Finean fragmentalism \citep{Fine2005,Lipman2020,Readfragment,CalosiForthcoming-CALFPA-3} raises the worry that the View from Everywhere's plurality of frame-relative realities is internally \textit{incoherent}. Yet each $g_{IJ}(\Phi)$ is frame-indexed \emph{by construction}, so apparent disagreements between $g_{IJ}(\Phi_r)$ and $g_{IJ}(\Phi_b)$ are not contradictions over a shared subject matter, but complete characterisations of \emph{distinct} physical situations, each individuated by its frame. Most importantly, Finean perspectivalism—of which fragmentalism is an option—differs from the physical perspectivalism discussed in this paper. 
Relationally constructed quantities via reference frames fall under what Fine calls \textit{internal} relativism, where quantities are internally indexed to a reference frame (e.g. ‘property p obtains within frame f’). By contrast, fragmentalism wants to vindicate the more controversial reality of \textit{unindexed} quantities—e.g. ‘property p obtains’—although placing them in different `fragments’ of reality, to avoid the joint obtaining of incoherent facts.
A comprehensive overview of all the terms used to refer to metaphysical positions such as fragmentalism, perspectivalism(s), external relativism and internal relativism in the context of GR is the subject of future work.}
If neutrality is instead understood as frame-independence, the connections represented by $\mathbf m$ would themselves count as neutral and the construction would no longer vindicate strong perspectivalism under this reading.

\section{Conclusion}\label{conclusion}

This paper has examined what physical situations relational observables in GR can be taken to represent. The fibre-bundle construction with two material reference frames (\S\S\ref{sec1}--\ref{VFN}) made explicit which gauge-invariant structures the \emph{View from Nowhere} and the \emph{View from Everywhere} share and where their ontological commitments diverge. The distinction between \emph{frame-independence} and \emph{frame-freedom} (\S\ref{terminology}) sharpened the interpretation of these structures.

The View from Nowhere locates ontological commitment also in the frame-free class $[m]$ alongside the family of relational observables.  In the setting with two material frames, a relational metric is Komar-complete only for the corresponding partial-projection equivalence class, while the whole-model class $[m]$ carries strictly more gauge-invariant content and is understood as the more comprehensive physical situation underlying the frame-relative descriptions. The View from Everywhere, by contrast, articulates a purely relational ontology: it accepts the difference in content, but takes each individual frame-relative situation to be ontologically comprehensive in it own right. 
The further gauge-invariant content is retained through the relational observables and their specified inter-frame maps. \emph{Relality} names this structured plurality without turning it into a further entity of which the individual situations are partial manifestations. The disagreement thus concerns how the reconstructed content is understood: as belonging to one shared physical situation or to distinct, connected situations exhausting the ontology.

This contrast suggests a question for the perspective-neutral approach to quantum reference frames \citep{Vanrietvelde2020,delaHametteperspective,Hhn2021}. Reading the perspective-neutral physical Hilbert space as a frame-free structure with ontological significance may yield a quantum version of the View from Nowhere. Whether the same formalism admits a View-from-Everywhere reading---which seems more aligned to \cite{Giacomini2019}'s original approach---also is to be investigated.\footnote{Private Correspondence.}

The compatibility of inter-perspectival connections with strong perspectivalism depends on the reading of perspective-neutrality. Under the frame-free reading, the GR construction showed that physically constrained connections between frame-relative observables do not require a shared neutral physical situation. Under the frame-independent reading, Adlam's reservations remain. The result concerns physical relations between frames; it does not establish the conditions for communication or shared knowledge. The same distinction may be useful in examining cross-perspective links in relational quantum mechanics \citep{Adlam2023,rovelliRQMnew,Adlam2026}. Whether classical frame-independent inter-frame translations can illuminate these quantum links requires a separate analysis of the relations between measurement outcomes, their records and the modal structure of RQM.

The metaphysical choice remains open, including the possibility of independent explanatory grounds for preferring the View from Nowhere. What the comparison establishes is that preserving physical content and connecting perspectives do not, within this formal construction, require a shared frame-free physical situation. The View from Everywhere can therefore be assessed as a formally valid alternative interpretation with explicit physical content. For philosophers and physicists alike, the framework provides a common basis for evaluating these alternatives: it makes clear what each interpretation retains, where their commitments diverge, and what further reasons would be needed to favour one over the other.

\section*{Acknowledgement}
I thank Henrique Gomes and Flaminia Giacomini for useful discussions.

\bibliography{BIB2.bib}

\end{document}